# Approximation Algorithms for Vehicle Routing Problems with Stochastic Demands on Trees

Shalabh Vidyarthi and Kaushal K Shukla
Department of Computer Engineering
Indian Institute of Technology (BHU), Varanasi
shalabh.vidyarthi.cse09@iitbhu.ac.in, kkshukla.cse@iitbhu.ac.in

**Abstract**

We consider the *vehicle routing problem with stochastic demands* (VRPSD) on tree structured networks with a single depot. The problem we are concerned with in this paper is to find a set of tours for the vehicle with minimum expected length. Every tour begins at the depot, visits a subset of customers and returns to the depot without violating the capacity constraint. Randomized approximation algorithm achieving approximation guarantees of 2 for *split-delivery* VRPSD, and 3 for *un-split delivery* VRPSD are obtained.

**Keywords**: Transportation, Vehicle Routing, Approximation Algorithm

## 1  Introduction

The capacitated vehicle routing problem arises when the customers are to be served by vehicles of limited capacity. The objective is to find set of routes each starting at the depot and visiting a subset of customers such that the value of the total distance travelled is minimized. The capacitated vehicle routing problem (VRP) is defined in [6] on a finite metric space ($V,d$), where $V$ is a finite set of locations/vertices and $d : V \times V \rightarrow R_+$ a distance function that is symmetric and satisfies the triangle inequality. There is a specified *depot* location $r \in V$, and the problem involves distributing (identical) items from the depot to other locations. Specifically, the depot $r$ has an infinite supply of items and a single vehicle of capacity $Q >= 0$ (initially located at the depot $r$) which is used to distribute the items. The demand $q_i \in \{0, 1... Q\}$ for each location $i \in V$. In the split delivery version of the problem the demand at a location can be satisfied by multiple visits. In the un-split version the demand at each location needs to be satisfied by a single visit.

In the Vehicle Routing Problem with Stochastic Demands, the demands involved are random variables with a known distribution in the range $\{0,1..., Q\}$. The exact value of the demand at any location is known only when the vehicle visits the location. The goal is to minimize the expected length. There are two variants of the said problem i.e. with split and un-split deliveries.

### Related Work

Throughout the text we consider the VRPSD on tree networks. A. Gupta et al[6] consider the said problem in general metrics and provide $(1+\alpha)$ approximation algorithm for split deliveries and $(2+\alpha)$ approximation algorithm for un-split deliveries. Where $\alpha$ ($\geq 1$) is the best approximation guarantee for the *Travelling Salesman Problem*. Bertsimas [3] considers the same problem providing the upper and lower bounds on the expectation values of the total length of the tours. A $(1+a+o(1))$-approximation algorithm is known for split-delivery VRPSD in the special case of *identical demand distributions* in the same paper.

Related to the tree metric version considered in this paper, Labbe et al. [1] and Karuno et al. [2] discuss some practical situations where tree shaped networks are encountered in VRPs. The problem Tree-CVRP is shown to be strongly NP-Complete by a reduction from the bin packing problem by Hamaguchi and Katoh[8].

### Prelimnaries

The following discussion holds for both split and un-split delivery versions of VRPSD. We may assume that none of the locations have a zero demand. Since the actual demands are revealed only when the vehicle visits a node, each node must be visited with a probability 1 In [6] a priori Travelling Salesman Tour is followed by the vehicle along with the return to the depots should the demands on a route exceed Q. However on tree networks, the minimum distance that can be covered in satisfying the demands at the various locations is 2S, where $S = \sum_{e \in E} e$. Let $D_i$ be the random variable representing the demand for customer i. Let $p_i(k) = Pr(D_i = k)$, $i \in \{1,..,n\}$ and $k \in \{1,..,Q\}$.

## 2 Bounds and Approximation Algorithms

**Lower Bound**

In this section an improved lower bound on the expectation value of the length of the VRPSD in trees is shown. Bertsimas [3] showed that:

$$E[R_{VRP}] \geq (2/Q)\sum_{r=1}^{n} d(0,r)E[Dr]. \quad (1)$$

Corresponding to the priori TSP tour in [6], the priori sequence for visiting the nodes is decided by a Depth First Search. Let the vertices be labeled [0, 1... n] in order of their visit in the DFS, where 0 represents the depot. Let $T_j$, represent the sub-tree visited between two visits to the depot and length of $T_j$ be $L_j$. The $T_j$ are not disjoint in the case of split deliveries. Let $w_j$ denote the node farthest from the depot visited in the tour j. Clearly

$$L_j \geq 2d(0, w_j) \quad (2)$$

Multiplying by the demand $i_r$ ($r \in T_j$) and taking summation over all nodes

$$L_j \sum_{r \in T_j} i(r) \geq 2\sum_{r \in T_j} d(0, w(j))i(r) \quad (3)$$

i(r) and w(j) are same as $i_r$ and $w_j$ respectively. Since on each tour $T_j$, $\sum i_r \leq Q$

$$L_j \geq (2/Q) \sum_{r \in T_j} d(0, w(j))i(r) \quad (4)$$

Taking summation over all the tours $T_j$

$$R_{TVRPSD} \geq (2/Q)\sum_{j=1}^{I} \sum_{r \in T_j} d(0, w(j))i(r) \quad (5)$$

where $I$ is the total number of tours required to fulfill the demands at all locations. Using $E[R_{VRP}] = \sum_{i1,...,in} p1(i1) .... pn(in)R(i1, i2,.., in)$ from [3]

$E[R_{TVRPSD}] \geq (2/Q) \sum p_1(i_1)....p_n(i_n)R(i_1,i_2,..,i_n)$

$$\sum_{j=1}^{I} \sum_{r \in T_j} d(0, w(j))i(r)$$
$$= (2/Q) \sum_{j=1}^{I} d(0, w(j)) \sum_{r \in T_j} E[Dr] \quad (6)$$

$E[R_{TVRPSD}] \geq 2/Q)\sum_{j=1}^{I} d(0, w(j)) \sum_{r \in T_j} E[Dr]$

## Approximation Algorithms

### Split Deliveries

In this section the split delivery problem is considered. The main idea used in [6] is initially filling the vehicle with a random number of items. Though it is contrary to a natural strategy of filling a vehicle to capacity, it yields a better expectation value. With slight modification in the SplitALG in [6] we achieve an approximation ratio of 2 for the split TVRPSD.

SplitALG for trees is as follows:

1) Compute the order in which the vertices are to be visited by applying a Depth First Search. Label the vertices [0, n], where 0 represents the depot. The vertices are visited in the order 1, 2.., n.
2) Choose a uniformly random value l in the range [0, Q] for initially loading the vehicle.
3) For each i = 1, …, n
   a) Let $Q_i$' items be present on the vehicle when it visits the customer i & $q_i$ be the demand.
   b) If $q_i < Q_i$', serve the demand at the node i and move to the node $q_{(i+1)}$.
   c) If $q_i = Q_i$', serve the demand at node i and return to the depot to refill to full capacity Q. Visit location (i+1) from the depot.
   d) If $q_i > Q_i$', serve $Q_i$' units at i and return to the depot for refill to full capacity Q. The vehicle serves Q- $q_i$ at i and proceeds to the (i+1)$^{th}$ vertex. $Q_{(i+1)}$' = Q – ($q_i$ - $Q_i$').

The initial load is uniformly random in the range [0, Q]. A node i is a breakpoint if the vehicle executes a refill trip from i to the depot. The node i is a breakpoint if $\sum_{j=1}^{i-1} q(i) < 1 + p.Q \leq \sum_{j=1}^{i} q(i)$. Since l is the only random variable with a uniform distribution in [0, Q],

Pr(i is a break-point) = $q_i/Q$

$S = \sum_{e \in E} e$, be the sum of all the edges in the tree network, is the length of the initial DFS-tour. The expected solution length is:

$$2S + (2/Q) \sum_{i \neq 0} q(i).d(0,i). \quad (7)$$

Taking expectation value over demands, the expected solution length is

$$2S + (2/Q) \sum_{i \neq 0} E[Dr].d(0,i).$$

2S is the length of the initially computed sequence.

$(2/Q)\sum_{i \neq 0} E[Dr].d(0,i) \leq (2/Q)\sum_{j=1}^{I} d(0, w(j)) \sum_{r \in T_j} E[Dr]$.

Therefore expected length of the solution is at most 2 times the optimal solution length for TVRPSD instance.

### Unsplit Deliveries

In this section the un-split variant of the problem is considered. The UnsplitALG is very similar to SplitALG for trees.

1) Compute the order in which the vertices are to be visited by applying a Depth First Search. Label the vertices [0, n], where 0 represents the depot. The vertices are visited in the order 1, 2.., n.
2) Choose a uniformly random value l in the range [0, Q] for initially loading the vehicle.
3) For each i = 1, …, n
   a) Let $U_i'$ items be present on the vehicle when it visits the customer i & $q_i$ be the demand.
   b) If $q_i < U_i'$, serve the demand at the node i and move to the node $q_{(i+1)}$.
   c) If $q_i = U_i'$, serve the demand at node i and return to the depot to refill to full capacity Q. Visit location (i+1) from the depot.
   d) If $q_i > U_i'$, two visits to the depot are required
      - In the first visit, the vehicle fills up $q_i$ units at r and serves the demand at i.
      - In the second visit, the vehicle fills up (Q + $U_i'$ - $q_i$) units at *r*, and returns to *i*.

It can be observed that the break points in SplitALG and UnsplitALG are identical. The probability that node i is the break-point is $q_i/Q$. The solution length in the above algorithm is 2S+(4/Q) $\sum_{i \neq 0} q(i).d(0,i)$. The expected solution length over demands is 2S + (4/Q) $\sum_{i \neq 0} E[Dr].d(0,i)$.

$$(2/Q)\sum_{i \neq 0} E[Dr].d(0,i) \leq \sum_{j=1}^{l} d(0,w(j))\sum_{r \in T_j} E[Dr]$$

The expected value of the additional distance to be travelled when the vehicle runs out of items is twice that of the split deliveries. Using the analysis similar to SplitALG, the expected solution length is at-most 3 times the optimal value for the TVRPSD instance.

## 3 Conclusions

In this paper we derived a lower bound for the expectation value of the length of SVRP tour over a tree network. We also presented constant factor approximation algorithms for the case of split and un-split deliveries. The open questions that arise is to improve these guarantees: Asano et. al [7] achieve an approximation guarantee of 1.35078 for the split delivery case when vehicle capacity is 1.